\documentstyle[12pt]{article}
\begin{document}
\newcommand{\bea}{\begin{eqnarray}}
\newcommand{\eea}{\end{eqnarray}}
\newcommand{\be}{\begin{equation}}
\newcommand{\ee}{\end{equation}}
\newcommand{\non}{\nonumber}
\newcommand{\th}{\theta}
\global\parskip 6pt
\begin{titlepage}
\begin{center}
{\Large\bf Subdivision Analysis of Topological}\\
\vskip .25in
{\Large\bf $\bf{Z_{p}}$ Lattice Gauge Theory}\\
\vskip .5in
Danny Birmingham \footnote{Supported by Stichting voor Fundamenteel
Onderzoek der Materie (FOM)\\
Email: Dannyb@phys.uva.nl}     \\
\vskip .10in
{\em Universiteit van Amsterdam, Instituut voor Theoretische Fysica,\\
1018 XE Amsterdam, The Netherlands} \\
\vskip .50in
Mark Rakowski\footnote{Email: Rakowski@tsmi19.sissa.it}   \\
\vskip .10in
{\em International School for Advanced Studies (SISSA/ISAS)\\
Via Beirut 2, 34014 Trieste, Italy}  \\
\end{center}
\vskip .10in
\begin{abstract}
We analyze the subdivision properties of certain lattice gauge theories
for the discrete abelian groups $Z_{p}$, in four dimensions.
In these particular models we show that the Boltzmann weights
are invariant under all $(k,l)$ subdivision moves, when the
coupling scale is a $p$th root of unity. For the case of manifolds
with boundary, we demonstrate analytically that Alexander type $2$ and
$3$ subdivision of a bounding simplex
is equivalent to the insertion of an operator which equals a
delta function on trivial bounding holonomies.
The four dimensional model then gives rise to an effective
gauge invariant three dimensional
model on its boundary, and we compute the combinatorially invariant
value of the partition function for the case of $S^{3}$ and
$S^{2}\times S^{1}$.
\end{abstract}
\vskip .5in
\begin{center}
ITFA-93-22/ SISSA-91/93/EP \\
June 1993
\end{center}
\end{titlepage}

\section{Introduction}
In \cite{top1,top2,top3}, a class of lattice gauge theories was introduced
which had appealing topological features. The program was essentially to study
lattice models which were based on a discrete Chern form type action; the
setting was a general simplicial complex which modeled a 4-manifold with
boundary. With the particular action functional that was adopted there,
it was possible to prove several remarkable subdivision properties when
the coupling parameter took on quantized values, for the gauge groups $Z_{2}$
and $Z_{3}$.
A complete analysis of the subdivision properties under moves which
subdivided the 4-dimensional simplex,
as well as for the 3-dimensional boundary,
was presented for these two models. It was found that the model was
subdivision invariant under all 4-dimensional moves which left the boundary
unchanged, while a subdivision of a bounding 3-simplex was shown to be
equivalent to the insertion of certain delta functions which trivialized
bounding holonomies.

   In this paper, we consider a variation of the model presented in
\cite{top1,top2,top3} which allows for a uniform extension to all finite
cyclic groups. The previous models discussed for $Z_{2}$ and $Z_{3}$
groups will appear as special cases of the general program. The change is
essentially to use angles to measure deviations from trivial holonomy in
roughly the same way that heat kernel methods are used as an alternative to
the usual Wilson action in Yang-Mills theory. We will also be able to
provide a complete analytic proof of the boundary subdivision properties
that were earlier proved by computer techniques in \cite{top3}.

   Given the subdivision properties of the bounding 3-dimensional simplex,
we will be able to identify a topological invariant as the continuum limit
of these models, and compute it for complexes which model $S^{3}$ and
$S^{2} \times S^{1}$. This model, which differs from the
naive partition function
by a scale factor associated with the number of 3-simplices in the bounding
complex, is related to structures recently considered in \cite{shap}.

\section{Action for ${\bf Z_{p}}$ Groups}
The models defined in \cite{top1,top2,top3} are based upon an action
which depends on two independent holonomies sharing a common vertex,
conveniently referred to as a bow-tie structure.
The action, evaluated on a $4$-simplex  $[0,1,2,3,4]$, was then
given by a sum
over permutations of the vertices, a typical term being of the form:
$(U-1)_{012} \;(U-1)_{034}$ for the $Z_{2}$ case, and
$(U -U^{-1})_{012} \;(U -U^{-1})_{034}$ for all other models.
Here, $U_{ijk}$ is the holonomy around the 2-simplex determined by
the vertices $i$, $j$, and $k$.
However, since $(U -1)$ and $(U - U^{-1})$ both
measure deviations from trivial holonomy, one
can consider simply replacing this by an angle while keeping the bow-tie
structure of the Boltzmann weight intact. This, in fact, allows one to
streamline the analysis for the general $Z_{p}$ case, and we confine
our attention to these actions in the following.

  The cyclic group $Z_{p}$ is represented multiplicatively by the $p$ roots of
unity $\exp[ 2 \pi i n/p]$, where $n$ is an integer in the set
$\{ 0, \cdots, p-1 \} $. We will take
the link variables of our theory to be this set of integers, and
the fundamental ``holonomy'' combination is defined by
\bea
n_{ijk} = n_{ij} + n_{jk} + n_{ki} \;\;\;\; {\em mod}\;\;\; p \;\; .
\eea
We also have the usual rule associated with a reversal of link orientation:
\bea
n_{ji} = - n_{ij} \;\;\;\; {\em mod}\;\;\; p\;\; .
\eea

  Our theory is defined by specifying the Boltzmann weight evaluated on
the 4-simplex $[0,1,2,3,4]$;
\be
W[0,1,2,3,4] = \exp [\beta S[0,1,2,3,4]]\;\;,  \label{bw}
\ee
where the action is given by
\bea
S[0,1,2,3,4] &=& (n_{012} \, n_{034} + n_{013} \, n_{042}
+ n_{014} \, n_{023} \non\\
&+& n_{102} \, n_{143} + n_{103} \, n_{124}
+ n_{104} \, n_{132}\non\\
&+& n_{201} \, n_{234} + n_{203} \, n_{241}
+ n_{204} \, n_{213}    \non\\
&+& n_{301} \, n_{342} + n_{302} \, n_{314}
+ n_{304} \, n_{321}\non\\
&+& n_{401} \, n_{423} + n_{402} \, n_{431}
+ n_{403} \, n_{412})\;\;,
\label{action}
\eea
and $\beta$ is the coupling parameter.
We are concerned in this paper with the behavior of the theory when
the scale factor $s = \exp[\beta]$ is a $p$th root of unity:
\bea
s = \exp[ 2 \, \pi\, i \, k /p] \;\; ,
\eea
for $k =\{0,1,\cdots , p-1\}$.
For later convenience,
we denote the Boltzmann weight for a given vertex ordering by:
\be
B[0,1,2,3,4] = s^{n_{012} \, n_{034}}\;\;.
\ee

\section{Behavior Under the $(k,l)$ Moves}
We  begin by considering the behavior of the theory under subdivision
moves of $(k,l)$ type. For closed manifolds
of dimension less than or equal to four,
these moves have been
shown to be equivalent to the Alexander type subdivisions \cite{alex,gross}.

{\bf Lemma:} The Boltzmann weights  of the $Z_{p}$ theory,
for a given vertex ordering, satisfy the conditions,
\be
B[0,1,2,3,4]\, B[0,1,2,4,5]\,B[0,1,2,5,3] = s^{n_{012} \, n_{345}}
\;\; ,\label{zbbb}
\ee
\be
B[0,1,2,3,4]\,B[1,2,0,4,3] = s^{n_{012} \, n_{014}}\,
s^{-n_{012}\, n_{013}}
\;\;,\label{zbb}
\ee
when the scale factor is a $p$th root of unity, $s^{p} =1$.

To prove this, one first notes that the relations are trivially satisfied
when $n_{012} =0$, so we can concentrate on the non-trivial values for
this holonomy.
In fact, it suffices to study only the case when $n_{012} =1$;
the remaining values lead to less restrictive constraints on the
scale parameter.
Since the holonomy angle $n_{ijk}$ is defined as the modular
sum of three link variables,  we immediately have the relation
\be
n_{034} + n_{045} + n_{053} - n_{345} = 0 \;\;\;\; {\em mod}\;\;\; p \;\;,
\ee
and the results follow.

This leads us to our first theorem:

{\bf Theorem 1:} The Boltzmann weights of the $Z_{p}$ model
satisfy the relation,
\bea
W[0,1,2,3,4]\,W[0,1,2,4,5]\,W[0,1,2,5,3]
&=&W[0,1,3,4,5]\,W[1,2,3,4,5]\non\\
&\,&W[2,0,3,4,5] \;\;,\label{WWW}
\eea
when $s^{p} =1$.

The proof of this is again straightforward, though a little tedious.
In verifying the result, one can use the relation
\bea
B[0,1,2,3,4] \, B[0,1,2,4,5]\, B[0,1,2,5,3]
&=&B[3,4,5,0,1] \, B[3,4,5,1,2]\non\\
&\,& B[3,4,5,2,0]\;\;,
\eea
which is a trivial consequence of (\ref{zbbb}), to eliminate all but
18 of the 90 terms. The identity (\ref{zbb}) then comes into play,
and the result is secured.

This theorem now allows us to analyze completely the remaining
$(k,l)$ moves, which we state as a corollary.

{\bf Corollary:} The Boltzmann weights of the $Z_{p}$ model
satisfy the following two relations:
\bea
W[0,1,2,3,4]\,W[0,1,2,5,3] &=& W[1,2,3,4,5]\, W[2,0,3,4,5]\,
W[0,1,3,4,5]\,\non\\
&\,&W[1,0,2,4,5]\;\;,
\eea
\bea
W[0,1,2,3,4]&=&W[5,1,2,3,4]\, W[0,5,2,3,4]\, W[0,1,5,3,4]
\,W[0,1,2,5,4]\non\\
&\,& W[0,1,2,3,5] \;\; ,
\eea
at the points $s^{p} =1$.

   The proof here is a simple application of Theorem 1, together
with the fact that,
\bea
W[0,1,2,3,4]^{-1} = W[0,1,2,4,3]
\eea
holds in this theory at the points $s^{p}=1$.

Having established invariance with respect to the $(k,l)$ moves, we can
state that the partition function of the $Z_{p}$ model is a
combinatorial invariant for all closed $4$-manifolds. However, these
identities also allow us to conclude triviality of the invariant for
all manifolds in this class.  In particular, we can show that these
four dimensional models actually reduce to behavior on the boundary.

To establish the triviality for closed $4$-manifolds, we consider the
identity of theorem $1$, in the form:
\begin{eqnarray}
1=& &W[0,1,2,3,4] \, W[0,1,2,4,5]\, W[0,1,2,5,3]\label{id} \\
& &W[1,0,3,4,5]\, W[2,1,3,4,5] \, W[0,2,3,4,5]\;\; .\nonumber
\end{eqnarray}
Written in this way, one
can recognize that the 4-simplices in this identity are actually the boundary
of a 5-simplex $[0,1,2,3,4,5]$;
\begin{eqnarray}
\partial\, [0,1,2,3,4,5] =
& &[1,2,3,4,5] - [0,2,3,4,5] + [0,1,3,4,5] -\\
& &[0,1,2,4,5] + [0,1,2,3,5] - [0,1,2,3,4]\;\; .\nonumber
\end{eqnarray}
If $t$ is any 5-simplex, we can write this compactly as:
\begin{eqnarray}
W[\partial \, t] = 1\;\; .
\end{eqnarray}

    Let $K$ denote a simplicial complex which models a 4-manifold, possibly
with boundary. For our purposes, the 4-simplices $\{ s_{i} \}$ in K are
most important,
\begin{eqnarray}
K = \sum_{i} \; s_{i} \;\; .
\end{eqnarray}
Now consider the abstract simplicial complex called the cone over $K$
\cite{JM},
which is obtained by adding a new vertex $x$ to the simplicial complex $K$, and
linking it to all other vertices; we denote this simplicial complex by
$x \ast K$. Computing the boundary of that complex, one sees
\begin{eqnarray}
\partial \, (x \ast K) = K - x \ast \partial K \;\; .
\end{eqnarray}
Given that the Boltzmann weight of the left hand side is just 1, we have then
\begin{eqnarray}
W[K] = W[ x \ast \partial K]\;\; ,\label{cone}
\end{eqnarray}
where we mean, more precisely, that
\begin{eqnarray}
W[K] = \prod_{i} \; W[s_{i}] \;\; .
\end{eqnarray}
The simplicity of equation (\ref{cone}) is striking; the implication
is that the Boltzmann weight of any four dimensional simplicial complex
$K$ is identical to that of the cone over its boundary. Essentially, the
cone construction is giving a canonical presentation - or framing - of
the boundary of $K$. Moreover, if $\partial K$ has several disjoint components
$M_{\alpha}$, then $K$ is a cobordism connecting them, and we immediately
have that
\begin{eqnarray}
Z[K] = \prod_{\alpha} \; Z[x \ast M_{\alpha}] \;\; .
\end{eqnarray}
This is one of the axioms for a topological field theory \cite{Atiyah}.

   Having established that we are dealing with a four dimensional gauge
theory which essentially reduces to something on the boundary, it is
natural to wonder about its interpretation as an intrinsically three
dimensional theory. As a gauge theory, we can gauge fix the links on any
maximal tree, and one such tree is given by the links which spew from
the vertex $x$. The value of the partition function is independent of how
we fix them, so we could always set those link variables to 1 say.
However one chooses to gauge fix these link variables, we can consider
the result to be a three dimensional lattice theory. If there is any
residual gauge invariance left, then it is not at all manifest, but
this is also reminiscent of the continuum Chern-Simons theories \cite{EWit}.

\section{Analysis of Boundary Subdivision}

   In this section, we will undertake an analysis of the theory when a
bounding simplex is subdivided by an Alexander move of type 2 or 3.

    Consider what happens to the Boltzmann weight of the theory when a
bounding 3-simplex $[0,1,2,3]$ is subdivided. Let $x$ be the cone vertex
discussed in the previous section, so that the partition function would
have the factor $W[0,1,2,3,x]$. Under subdivision where we add a new
vertex $c$ to the center of the tetrahedron, we would then consider a
new set of Boltzmann weights which are unchanged except that we would
replace the factor $W[0,1,2,3,x]$ by the quantity,
\begin{eqnarray}
W[c,1,2,3,x]\; W[0,c,2,3,x]\; W[0,1,c,3,x]\; W[0,1,2,c,x] \label{sdw}
\end{eqnarray}
in the partition function, and sum over the new link variables.
However, the main identity that we established above,
namely (\ref{WWW}), says that this product of four weights is
equal to
\begin{eqnarray}
W[0,1,2,3,x]\; W^{-1}[0,1,2,3,c] \;\; .
\end{eqnarray}
Here, $W^{-1}$ denotes the inverse value which, for the Boltzmann weights
in our construction, is equivalent to simply an odd permutation of the
vertices; $W^{-1}[0,1,2,3,4] = W[0,1,2,4,3]$.
We see then that the subdivided Boltzmann weights represented by (\ref{sdw})
are precisely equivalent to having introduced an extra factor
$W^{-1}[0,1,2,3,c]$ into the original assembly of Boltzmann weights. It
is then crucial to understand how the theory behaves under insertions
of the kind:
\begin{eqnarray}
I[0,1,2,3] = \frac{1}{|G|^{4}}\; \sum_{n_{ci}} \;
W^{-1}[0,1,2,3,c]\;\; ,\label{insert}
\end{eqnarray}
where the sum is over the four link variables connected to $c$,
and $|G|$ is the order of the gauge group. At the
trivial points where (\ref{WWW}) holds, namely $s=1$, this
quantity is manifestly 1. We are interested in investigating the
nontrivial roots of unity.
Let $\delta (n)$ denote the Mod p delta function which
is $1$ for $n =0$ mod p, and zero otherwise.

{\bf Theorem 2:} The insertion $I[0,1,2,3]$ is equal to:
\begin{eqnarray}
\delta(n_{012})\; \delta(n_{013})\;
\delta(n_{023}) \;\delta(n_{123})
\end{eqnarray}
at the non-trivial roots of unity, $s^{p} =1$, in the $Z_{p}$ model.

In order to establish this result, we require the formula
\be
\frac{1}{|G|}  \sum_{k=0}^{p-1} \exp[2\pi i k n/p] = \delta(n)  \;\;.
\label{sum}
\ee
One can now sum over the four link variables in succession to yield
the four delta functions.
In addition, one  finds that the remaining portion of the Boltzmann
weight reduces to $1$, upon implementation of the delta
function constraints.

Thus, the insertion is 1 if all holonomies on the bounding 3-simplex
are trivial, and zero otherwise.

It is also interesting to consider Alexander type 2 subdivision of a 2-simplex
which belongs to the bounding 3-manifold. Since the bounding space is
a manifold, a given 2-simplex, say $[0,1,2]$, will be shared by precisely
two 3-simplices; we denote their sum by $[0,1,2,3] - [0,1,2,4]$. The
requirement of the relative minus sign is dictated by the fact that we
have a 3-manifold without boundary. Under type 2 subdivision, we
add a new vertex $c$ to the center of the $[0,1,2]$ face, and link it
to the other vertices,
\begin{eqnarray}
[0,1,2] \rightarrow [c,1,2] + [0,c,2] + [0,1,c] \;\; .\label{t2}
\end{eqnarray}
Now, in the Boltzmann weights appropriate to the original complex, one
will find the product,
\begin{eqnarray}
W[0,1,2,3,x] \; W^{-1}[0,1,2,4,x]\;\; .
\end{eqnarray}
In the subdivided situation, each of these two factors will be replaced
by a product of three Boltzmann weights according to the structure
of (\ref{t2}). If we again use the identity (\ref{WWW}), one finds that
the subdivided situation is equivalent to the insertion of the following
factor in the original product of Boltzmann weights:
\begin{eqnarray}
I'[0,1,2,3,4] = \frac{1}{|G|^{5}}\; \sum_{n_{ci}}\;
W^{-1}[0,1,2,3,c] \; W[0,1,2,4,c] \;\; .
\end{eqnarray}

Let $n_{ijkl}= n_{ij}+n_{jk}+n_{kl}+n_{li}$ denote the
holonomy angle through four vertices;
we then have the following result.

{\bf Theorem 3:} The quantity $I'[0,1,2,3,4]$ is equal to,
\begin{eqnarray}
\delta(n_{012})\; \delta(n_{0314})\;
\delta(n_{1324})\;\delta(n_{2304})\;\; ,
\end{eqnarray}
at the same points as in Theorem $2$.

Again, the proof here requires only a straightforward application
of the summation rule (\ref{sum}). As before, the remainder of the
Boltzmann weight reduces to $1$ after the delta function constraints
are applied.

Notice that there is one 3-vertex holonomy around the 2-simplex $[0,1,2]$
which is the face common to the two 3-simplices that have been glued
together; the other 4-vertex holonomies are just products of the more
elementary holonomies. Since any 2-simplex on the boundary is common
to precisely two bounding 3-simplices, the restriction imposed by type
2 subdivision is actually equivalent to that from the type 3 move in the
full partition function.

The above relations allow us to obtain the continuum limit
of these models in an economical fashion. Starting from any finite simplicial
complex, we simply perform a single Alexander subdivision of type $3$
on each of the bounding $3$-simplices. The Boltzmann weight is then
given solely by an assembly of delta functions. In fact, since these
insertions are themselves gauge invariant, we have distilled a gauge
invariant model from the original four dimensional theory.
Actually, in order to ensure that the value of the partition function
is invariant under further subdivision,  a scaling of the original
partition function must be performed.
Defining the quantity
\be
J[0,1,2,3] =
\frac{1}{|G|} \sum_{n_{ci}} \delta(n_{c01})
\delta ( n_{c02}) \delta (n_{c03}) \delta ( n_{c12})
\delta (n_{c13}) \delta (n_{c23})\;\;,
\ee
and performing the integrations with the aid of the formula
\be
\sum_{k=0}^{p-1} \delta (k) = 1\;\;,
\ee
we obtain the result:
\be
J[0,1,2,3]=
\delta (n_{012}) \delta (n_{013})
\delta (n_{023}) \delta (n_{123})     \;\;.
\ee
Similarly, the quantity
\bea
J'[0,1,2,3,4] &=&
\frac{1}{|G|} \sum_{n_{ci}}
\delta (n_{c01}) \delta (n_{c02}) \delta (n_{c03})
\delta (n_{c12}) \delta (n_{c13}) \delta (n_{c23})\non\\
&.&\delta (n_{c04}) \delta (n_{c14}) \delta (n_{c24})   \;\;,
\eea
can easily be shown to be:
\be
J'[0,1,2,3,4] = \delta (n_{012}) \delta (n_{013}) \delta (n_{023})
\delta (n_{123})\delta ( n_{014}) \delta ( n_{024})
\delta ( n_{124})   \;\;.
\ee
The quantities $J$ and $J'$ represent an assembly of delta functions
after Alexander subdivision of type $3$ and $2$.
Since the number of $3$-simplices
increases by $3$ and $4$ under these moves, it is a simple exercise to
conclude that the partition function
\be
Z = |G|^{(N_{3}(\partial K) - N_{1})} \sum_{n_{ij}} W[K]\;\;,\label{sdip}
\ee
is subdivision invariant, where $W[K]$ is a Boltzmann weight trivializing
all holonomies on the bounding 3-simplex.  Here, $N_{3}(\partial K)$
is the number of
bounding $3$-simplices, and $N_{1}$ is the number of link variables in
$K$.

\section{Computations}

Having established the fact that one can extract a subdivision invariant
model from the four dimensional setting, we now turn to some explicit
examples. In particular, we compute the partition function for
the case of a four manifold with $S^{3}$ or $S^{2} \times S^{1}$ boundary.

Consider first the case of $S^{3}$.
As we have seen, one can gauge fix the link variables which spew from
the cone vertex, leaving an effective theory on the boundary which is
itself gauge invariant. We can thus confine our attention to a simplicial
complex for $S^{3}$. A suitable choice is given by the boundary of
the $4$-simplex $[0,1,2,3,4]$ as follows:
\be
K = [0,1,2,3] - [0,1,2,4] + [0,1,3,4] -[0,2,3,4] + [1,2,3,4]\;\;.
\label{s3}
\ee

In order to compute the partition function, one can gauge the links
on a maximal tree, an example being:
\be
n_{01} = n_{12} = n_{23} = n_{34} = 0\;\;.
\ee

Corresponding to the simplicial complex (\ref{s3}), we have ten holonomy
constraints; however, the Bianchi identity can be used to eliminate
all but six of these. The remaining set can now be analyzed in the
presence of the gauge fixing, and one sees that all
$10$ link variables assume a value of $0$.
The subdivision invariant value for the partition function (\ref{sdip})
is then given by:
\bea
Z[S^{3}] = \frac{1}{|G|} \;\;.
\eea

Turning now to the case of a $4$-manifold with $S^{2} \times S^{1}$
boundary, we have the following simplicial complex for $S^{2} \times S^{1}$:
\bea
K &=& [0,1,2,4]\,\,\, - \,\,\, [1,2,3,5]
\,\,\,-\,\,\, [0,1,3,4]\,\, + \,\,\, [0,2,3,4]    \non\\
&+& [1,2,4,5] \,\,\,- \,\,\, [2,3,5,6] \,\,\,
-\,\,\, [1,3,4,5]\,\, + \,\,\,[2,3,4,6]    \non\\
&+& [2,4,5,6]\,\,\, - \,\,\,[3,5,6,7] \,\,\,
- \,\,\,[3,4,5,7]\,\, + \,\,\,[3,4,6,7]    \non\\
&+& [4,5,6,4']\,\, - \,\,[5,6,7,5'] \,\,
- \,\,[4,5,7,4']\, + \,\,[4,6,7,4']      \non\\
&+& [5,6,4',5']\, - \,[6,7,5',6']\,
- \,[5,7,4',5'] + \,[6,7,4',6']        \non\\
&+& [6,4',5',6'] - [7,5',6',7']
- [7,4',5',7'] + [7,4',6',7']          \non\\
&+& [4',5',6',0] - [5',6',7',1]
- [4',5',7',0] + [4',6',7',0]          \non\\
&+& [5',6',0,1]\, - \,[6',7',1,2] \,
- \,[5',7',0,1] + \,[6',7',0,2]        \non\\
&+& [6',0,1,2]\,\, - \,\,[7',1,2,3] \,\,
-\,\, [7',0,1,3]\, + \,\,[7',0,2,3]\;\;.
\label{s2xs1}
\eea
It requires a little more work to verify that this is indeed a
suitable simplicial complex, but it can be obtained by first constructing a
complex for $S^{2} \times I$, where $I$ is the unit interval.
The two $S^{2}$ boundaries are then identified, yielding (\ref{s2xs1}).

In this case, one has $48$ link variables, and a maximal tree involves
the gauge fixing of $11$ of these, to $0$ say. As before, the Bianchi
identity is used to obtain the independent holonomy constraints,
and upon implementation of the gauge fixing conditions, these can be
resolved. In fact, one finds a single constraint which specifies that $10$
of the link variables are equal,
with the remaining ones being set to zero.
The value of the partition function (\ref{sdip}) is therefore
\bea
Z[S^{2} \times S^{1}] = 1 \;\; .
\eea
We should note that these are equal to the values obtained in the models
presented in \cite{DW}; however, the value of the partition function
on $RP^{3}$ (which was 0 in those models) must necessarily be different.

\section{Concluding Remarks}

As we have seen from our analysis of boundary subdivision, the
original four dimensional theory spawns a gauge invariant model on
its boundary. It would be interesting to determine to what extent these
results can be extended to the nonabelian situation; one might expect
the quantum group case to give some insight in this regard. We would also like
to examine the duality properties of these models,
and to explore more fully the relationship with those considered in
\cite{DW,Alt}.


\begin{thebibliography}{99}
\bibitem{top1} D. Birmingham and M. Rakowski, {\em Subdivision Invariant
Models in Lattice Gauge Theory}, YCTP-P4-93 preprint, February 1993.
\bibitem{top2} D. Birmingham and M. Rakowski, {\em Combinatorial
Invariants from Four Dimensional Lattice Models}, YCTP-P6-93 preprint,
March 1993.
\bibitem{top3} D. Birmingham and M. Rakowski, {\em Combinatorial
Invariants from Four Dimensional Lattice Models: II}, YCTP-P11-93 preprint,
May 1993.
\bibitem{shap} S. Chung, M. Fukuma and A. Shapere, {\em Structure of
Topological Lattice Field Theories in Three Dimensions}, Cornell
preprint, CLNS 93/ 1200, May 1993.
\bibitem{alex} J.W. Alexander, {\em The Combinatorial Theory of
Complexes}, Ann. Math. 31 (1930) 292.
\bibitem{gross} M. Gross and S. Varsted, {\em Elementary Moves and
Ergodicity in $d$-Dimensional Simplicial Quantum Gravity}, Nucl. Phys.
B378 (1992) 367.
\bibitem{JM} J. Munkres, {\em Elements of Algebraic Topology},
Addison-Wesley, Menlo Park, 1984.
\bibitem{Atiyah} M.F. Atiyah, {\em Topological Quantum Field Theories},
Publ. Math. IHES 68 (1988) 175.
\bibitem{EWit} E. Witten, {\em Quantum Field Theory and the Jones
Polynomial}, Commun. Math. Phys. 121 (1989) 351.
\bibitem{DW} R. Dijkgraaf and E. Witten, {\em Topological Gauge Theories
and Group Cohomology}, Commun. Math. Phys. 129 (1990) 393.
\bibitem{Alt} D. Altschuler and A. Coste, {\em Quasi-Quantum Groups,
Knots, Three-Manifolds, and Topological Field Theory},
Commun. Math. Phys. 150 (1992) 83.
\end{thebibliography}
\end{document}